\documentclass[pra,twocolumn,amsmath,amssymb,superscriptaddress,showpacs]{revtex4}
\usepackage{graphics}
\usepackage{epsfig}
\newcommand{\bra}[1]{\ensuremath{\left\langle{#1}\right\vert}}
\newcommand{\braket}[1]{\ensuremath{\left\langle{#1}\right\rangle}}
\newcommand{\ket}[1]{\ensuremath{\left|{#1}\right\rangle}}
\newcommand{\ad}{\ensuremath{a^\dagger}}
\def\be{\begin{equation}}
\def\ee{\end{equation}}
\def\eea{\end{eqnarray}}
\def\bea{\begin{eqnarray}}

\newcommand{\va}[1]{\ensuremath{(\Delta#1)^2}}

\newcommand{\ex}[1]{\ensuremath{\langle{#1}\rangle}}

\begin{document}
\title{Number operator-annihilation operator uncertainty as an alternative
of the number-phase uncertainty relation}
\date{\today}
\begin{abstract}
We consider a number operator-annihilation operator uncertainty
as a well behaved alternative to the number-phase uncertainty relation,
and examine its properties. We find a formulation in which the bound
on the product of uncertainties
depends on the expectation value of the particle number.
Thus, while the bound is not a constant,
it is a quantity that can easily be controlled
in many systems. The uncertainty relation is approximately
saturated by number-phase intelligent states.
This allows us to define
amplitude squeezing, connecting coherent states to Fock states,
without a reference to a phase operator.
We propose several setups for an experimental verification.
\end{abstract}

\author{I\~nigo Urizar-Lanz}
\affiliation{Department of Theoretical Physics, The University of the Basque Country, P.O. Box 644, E-48080
Bilbao, Spain}
\author{G\'eza T\'oth}
\email{toth@alumni.nd.edu}
\affiliation{Department of Theoretical Physics, The University of the Basque Country, P.O. Box 644, E-48080
Bilbao, Spain}
\affiliation{IKERBASQUE, Basque Foundation for
Science,
E-48011 Bilbao, Spain}
\affiliation{Research Institute for Solid State Physics and Optics,
Hungarian Academy of Sciences, \\ P.O. Box 49, H-1525 Budapest,
Hungary}

%
%


\maketitle

\section {Introduction}

Finding a phase operator
conjugate to the number operator and constructing number-phase uncertainty relations has
an extensive literature \cite{C68,L95,BR97}. However,
defining a Hermitian phase operator for an infinite system, i.e.,
a harmonic oscillator, is not possible,
and all the different approaches must make certain compromises.

Historically, the first important contribution was that of
Dirac, who introduced the phase observable $\phi$
based on the decomposition of the annihilation operator as $a=R\exp(i\phi)$  \cite{D27}.
Assuming that $R$ and $\phi$
are Hermitian operators,
one obtains
$R=N^{1/2},$ where $N=\ad a$ is the number
operator. Hence, $\exp(i\phi)$ must be equal to
\be
E=\sum_{n=0}^{\infty}\ket{n}\bra{n+1}=(N+1)^{-1/2}a,
\label{E}
\ee
However, $E$ is not unitary, thus $\phi$ cannot be Hermitian either.

Several methods have been presented to circumvent the difficulties above.
Since there are extensive reviews on the topic  \cite{C68,L95,BR97}, we cite
only the literature that is directly connected to our approach.
Susskind and Glogower \cite{SG64} constructed the
Hermitain operators $C=\tfrac{1}{2}(E+E^\dagger)$ and
$S=\tfrac{1}{2i}(E-E^\dagger)$ to describe the quantum phase.
They obtained
uncertainty relations with them, however,
for the description of the phase two
operators were needed.
In order to overcome this inconvenience,
L\'evy-Leblond \cite{L76} suggested the use of
the non-Hermitian $E$ defined in Eq.~(\ref{E}).
He argued that physical quantities could also be
represented by non-Hermitian operators,
interpreted the meaning of variance for such operators
and wrote down uncertainty relations with $N$ and $E.$
Later, Hermitian phase operators were constructed for
finite systems \cite{BR97,PB88}. This makes it possible to
carry out a calculation for an expression with the phase operator
for a finite dimension $D,$
and then take the limit $D\rightarrow\infty,$ which
provides the value corresponding to the infinite dimensional case.
By use of this theoretical background, the number-phase uncertainty relation could
be obtained and the states saturating them, called
number-phase intelligent states, were identified \cite{VP90}.
The procedure that provides a connection from one number-phase
intelligent state to another one with a smaller number variance
is called amplitude squeezing \cite{AM87}.

Connected to these ideas, in this paper we choose the two operators to be
not $N$ and $E=(N+1)^{-1/2}a$, but simply
$N$ and the annihilation operator $a$.
We present the relation
\be
\bigg[\va{N}+\frac{1}{4}\bigg]\bigg[\va{a}+\frac{1}{2}\bigg]
\ge \frac{\ex{N}}{4}+\frac{1}{8}.
\label{unc3}
\ee
We will show that states saturating
the number-phase uncertainty are very close to saturating Eq.~(\ref{unc3}).
This makes it possible to define
amplitude squeezing without a reference to a phase operator.

In addition to its connections to quantum optics, this problem is
also interesting from the point of view of quantum information theory.
A family of uncertainty relations with $N$ and $a$ has already appeared
in Ref.~\cite{TC03}, and has been
used for the detection of quantum entanglement \cite{uncert,entag,entdet}.
Such uncertainty relations made it possible to construct
entanglement conditions with small experimental requirements.
Remarkably, these conditions detect non-Gaussian entangled states
that cannot be detected based on the first and second moments of the
quadrature components \cite{corrmat}.
The uncertainty relation Eq.~(\ref{unc3}) presented in this paper
can be seen as a single relation replacing the family of
uncertainty relations described in Ref.~\cite{TC03}.
For given $\ex{N},$ Eq.~(\ref{unc3})
identifies most of the values for the variances of $N$ and $a$
that are not allowed by quantum physics \cite{H87}.

The paper is organized as follows. In Sec.~II, we discuss how the variance
of the annihilation operator can be defined. In Sec.~III, we derive
the uncertainty relation Eq.~(\ref{unc3}).
In Sec.~IV, we discuss
the tightness of the uncertainty relation presented.
Finally, in Sec.~V we discuss
possible physical tests of the proposed uncertainty relation.
In the Appendix we present uncertainty relations
for two-mode systems.

\section {Variance of the annihilation operator }

In this section, we will discuss
the definition and the properties of the variance of the
annihilation operator. We will relate it to quadrature
independent properties of the quantum state.

We define the variance of a non-Hermitian operator $A$
as \cite{L76,remark}
\be
\va{A}=\ex{A^\dagger A}-\ex{A^\dagger}\ex{A}.
\label{vaa}
\ee
Note that, for non-Hermitian operators usually we have
 $\va{A} \ne \va{A^\dagger}.$

Let us now consider the $A=a$ case.  $\va{a}$ is zero {\it only} for
coherent states. The variance $\va{a}$ measures, in a sense, how close the quantum state is
to a coherent state. For this reason, it has been used to study the dynamics of various quantum
systems  (e.g., see Refs.~\cite{SG93, KO99,SM06,HK94}).

Let us now interpret $\va{a}$ by relating it to the quadrature components.

(i) Let us define the quadrature components as
\bea
x_\beta&=&\frac{ae^{+i\beta}+\ad e^{-i\beta}} {\sqrt{2}},\nonumber\\
p_\beta&=&\frac{ae^{+i\beta}-\ad e^{-i\beta}} {\sqrt{2}i},
\label{quad}
\eea
where $\beta$ is real. Then, one finds that
\be
\va{a}=\frac{\va{x_\beta}+\va{p_\beta}}{2}-\frac{1}{2}.
\label{sumx2p2}
\ee
Hence, $\va{x_\beta}+\va{p_\beta}$ is independent of the choice of the angle $\beta$ \cite{S86}.

When discussing the invariance properties of $\va{a},$ it is instructive to point out its connection
to the correlation matrix defined as
\begin{eqnarray}
\Gamma_\beta=\;\;\;\;\;\;\;\;\;\;\;\;\;\;\;\;\;\;\;\;\;\;\;\;
\;\;\;\;\;\;\;\;\;\;\;\;\;\;\;\;\;\;\;\;\;\;\;\;\;\;\;\;\;\;\;
\;\;\;\;\;\;\;\;\;\;\;\;\;\;\;\;\;\;\;\;\;\;\;\;\;\nonumber\\
\left(
\begin{array}{cc}
\va{x_\beta} & \frac{1}{2}(\langle \Delta x_\beta \Delta p_\beta+ \Delta p_\beta \Delta x_\beta\rangle) \\
\frac{1}{2}(\langle \Delta x_\beta \Delta p_\beta+ \Delta p_\beta \Delta x_\beta\rangle) & \va{p_\beta} \\
\end{array}
\right).\nonumber\\
\end{eqnarray}
One can obtain $\Gamma_{\beta'}$ from $\Gamma_{\beta}$ through orthogonal transformations.
However, the trace of $\Gamma,$ which equals to $\va{x_\beta}+\va{p_\beta},$
remains invariant under such transformations.

Thus, since $\va{x_\beta}+\va{p_\beta}$ is independent of
$\beta,$ it seems to be a good measure of
the uncertainty
of the orthogonal quadrature components.
Note that an alternative measure could be the product $\va{x_\beta}\va{p_\beta},$
however, it is not independent from $\beta.$

(ii) In another context, $\va{a}$ can be expressed
as
\be
\va{a}=\frac{1}{2\pi}\int_{\tilde{\beta}=0}^{2\pi}{\va{x_{\tilde{\beta}}}d\tilde{\beta}}-\frac{1}{2}.
\label{sumx2p2b}
\ee
Thus, $\va{a}$ is connected to the average variance of the quadrature components
$x_\beta.$ That is, if $\beta$ is chosen randomly between $0$ and $2\pi$
according to a uniform probability distribution then $\va{a}+\frac{1}{2}$ gives the expectation value
of the quadrature variance $\va{x_\beta}.$

(iii) Finally, let us examine the connection between $\va{a}$ and important properties of the Wigner function of the quantum state. For the following discussion, as well as in the rest of the paper, we will leave the  $\beta$
subscript, and will use $x$ and $p$ in the sense of $x_{0}$ and $p_{0},$
respectively.  $\va{a}$ gives information on the sharpness of the peak
of the Wigner function $W(x,p)$ of the state  since \cite{GZ04}
\bea
&&\va{x}+\va{p}\nonumber\\
&&\;\;\;\;\;\;\;\;\;\;\;\;\;\;\;\;=\int \left[ (x-\ex{x})^2+
(p-\ex{p})^2 \right]  W(x,p) dx dp.\nonumber\\\label{ww}
\eea
For states with a non-negative Wigner function (i.e., squeezed coherent states),
$2\va{a}+1$ is the sum of the squared widths of the Wigner function
in two orthogonal directions.
The sharpest peak is obtained for the coherent states for which Eq.~(\ref{ww})
is the smallest.

\section {Uncertainty relation with the number and the annihilation operators }

In this section, we will present a simple derivation of Eq.~(\ref{unc3})
and relate it to the uncertainty relation with the variances of $N$ and $E.$
We will also discuss how to improve the relation Eq.~(\ref{unc3}).

We start from the two Heisenberg uncertainty relations
\bea
\va{N}\va{p}&\ge&\tfrac{1}{4}|\ex{x}|^2,\nonumber\\
\va{N}\va{x}&\ge&\tfrac{1}{4}|\ex{p}|^2,\label{Heisenberg}
\eea
where we used the fact that for operators $A$ and $B$ we have
$\va{A}\va{B}\ge \tfrac{1}{4}|\ex{[A,B]}|^2$ \cite{GZ04}.
Summing the two inequalities of Eq.~(\ref{Heisenberg}),
using Eq.~(\ref{sumx2p2}) and
 $|\ex{p}|^2+|\ex{x}|^2=2|\ex{a}|^2$,
one obtains the following uncertainty relation with $N$ and
$a$
\be
\va{N}\bigg[\va{a}+\frac{1}{2}\bigg]\ge \frac{1}{4} |\ex{a}|^2.
\label{unc}
\ee
From Eq.~(\ref{unc}) it follows that knowing $\ex{a},$ which determines the "center" of the Wigner function $W(x,p)$ of the state, and $\va{a},$ which is based on the the width of the Wigner function
in two orthogonal directions, a lower bound for the particle number fluctuation can be obtained.

The bound in the uncertainty relation Eq.~(\ref{unc}) is not a constant:
it depends on $\ex{a},$ which is zero for a wide class of states.
It would be meaningful to find a similar relation with a constant bound
or at least with a bound depending on a quantity
that is easily measurable and controllable.

We will now construct a relation in which the bound depends on
$\ex{N}$ rather than on $\ex{a}.$ For that, we
add $[\va{a}+\frac{1}{2}]/4$ to both sides of Eq.~(\ref{unc}),
and using $\va{a}=\ex{N}-|\ex{a}|^2,$
we obtain Eq.~(\ref{unc3}).
The right hand side of Eq.~(\ref{unc3}) is minimal
for the vacuum
$\ket{0}$. In all other cases the right hand side
is greater than $\tfrac{1}{8}$, thus the uncertainty
finds some part of the $\va{a}$--$\va{N}$ plane
inaccessible for quantum states.

Next, we will relate Eq.~(\ref{unc3}) to the uncertainty relation
with $N$ and $E.$ For that, we determine the form of Eq.~(\ref{unc3}) for the case of
large $N$ and $\va{N}\ll \ex{N}^2.$ In this case $a\approx \sqrt{\ex{N}}E$ and we obtain
\be
\va{N}\va{E}\gtrsim \frac{1}{4}\left[1-\va{E}\right],
\ee
which is in accordance with the results of Ref.~\cite{L76}
\be
\va{N}\va{E} \ge \frac{1}{4}\left[1-\va{E}-\langle P^{(0)}\rangle\right],
\ee
where $P^{(0)}=\ket{0}\bra{0}.$

Finally, Eq.~(\ref{unc3}) can be improved by use of the Robertson-Schr\"odinger inequalities \cite{GZ04}.
First, they can be used to improve the two uncertainty relations in Eq.~(\ref{Heisenberg}).
 Then, after steps similar to the previous ones, we obtain
\be
\bigg[\va{N}+\frac{1}{4}\bigg]\bigg[\va{a}+\frac{1}{2}\bigg]
\ge \frac{\ex{N}}{4}+\frac{1}{8}+
\frac{1}{4}\vert \ex{\{\Delta N,\Delta a\}_+}\vert^2,
\label{unc4}
\ee
where $\{A,B\}_+=AB+BA$ is the anticommutator.

\section {Tightness of the inequality}

\begin{figure}
\centerline{\epsfxsize=3.in\epsffile{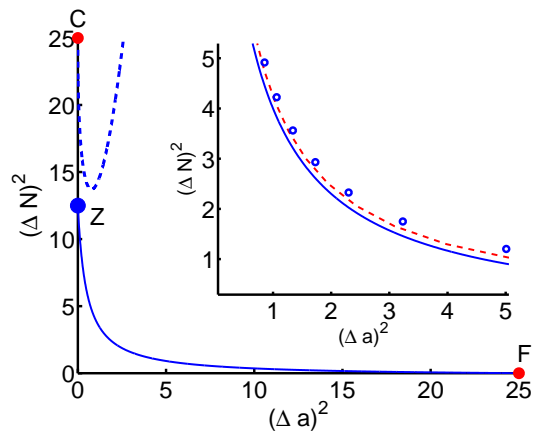}}
\caption{(Color
 online) Numerical test of the inequality (\ref{unc3}).
F refers to the Fock state $\ket{n=25},$
C refers to the coherent state $\ket{\alpha=5},$
Z refers to the point that saturates inequality (\ref{unc3}) for $\va{a}=0.$
(solid) Boundary of the region defined by Eq.~(\ref{unc3}) for $\ex{N}=25$.
All points below this line correspond to aphysical $\va{a}$--$\va{N}$
values.
(dashed) Points corresponding to squeezed coherent states. The equation of the curve is given in Eq.~(\ref{scs}).
Inset: (solid) Boundary of the region defined by Eq.~(\ref{unc3}).
(dashed) Points corresponding to states with a Gaussian wave vector Eq.~(\ref{Gauss}).
(circles) States corresponding to photon added coherent states Eq.~(\ref{pac}).
}
\label{fig_region}
\end{figure}

In this section, we will investigate the tightness of
Eq.~(\ref{unc3}), and will look
for quantum states that are close to
saturating it. We will also discuss that
the states saturating the left hand side of Eq.~(\ref{unc3})
interpolate between coherent states and Fock states.

Our inequality does not contain
the highest possible lower bound. The reason for that is
that we constructed Eq.~(\ref{unc3}) by summing the two
uncertainty relations in Eq.~(\ref{Heisenberg}). While the
new relation Eq.~(\ref{unc}) is valid, it is not tight since the
two uncertainty relations in Eq.~(\ref{Heisenberg})
are saturated by different states.
Thus, the tightness of the bound in Eq.~(\ref{unc3}) must be
verified.

In Fig.~1, we plotted the points corresponding to values of $[\va{a},\va{N}]$
that saturate
Eq.~(\ref{unc3}) for $\ex{N}=25.$
All points below this line violate the relation Eq.~(\ref{unc3}).
For Fock states,
\be \va{a}_{\rm Fock}=\ex{N},\;\;\;\; \;\;\;\;\;\;
\va{N}_{\rm Fock}=0.\ee
Hence, Fock states saturate Eq.~(\ref{unc3}).
For coherent states we have \be\va{a}_{\rm coh}=0,\;\;\;\; \;\;\;\;\;\;
\va{N}_{\rm coh}=\ex{N}.\ee For $\va{a}=0,$
the particle number variance saturating Eq.~(\ref{unc3})
is $\va{N}=\tfrac{1}{2}\ex{N}.$
This already shows that the lower bound in Eq.~(\ref{unc3})
cannot be optimal because for $\va{a}=0$
there is no quantum state with a smaller particle number variance
than $\ex{N}.$
States minimizing $\va{N}$ for $0<\va{a}<\ex{N}$, in a sense, interpolate between
coherent states and Fock states.

We will now examine the tightness of Eq.~(\ref{unc3}) numerically through
choice of appropriate trial states. Our search can be simplified by noting that
it is sufficient to search over  wave vectors with non-negative real elements.
To see this, let us  consider a state of the form $\ket{\Psi}=\sum_n \vert c_n \vert e^{i\phi_n} \ket{n}.$
One finds that if all angles $\phi_n$ are set to zero,
 $\va{N}$ and $\ex{N}$ do not change. On the other hand, $\vert \ex{a} \vert$ cannot decrease.
 Hence, $\va{a}=\ex{N}-\vert \ex{a} \vert^2$ cannot increase.
Thus, it is sufficient to search over states with all $\phi_n=0.$
Moreover, a state with $\phi_n={\rm const.}\times n$ will give the same values
for $\va{N}, \ex{N}$ and $\va{a}$ as does a state with $\phi_n=0.$ 

\subsection{Gaussian wave vector}

Let us consider states with a Gaussian state vector
\be
\ket{N_0,\Delta}=\frac{1}{C}\sum_n \exp\left[-\frac{(n-N_0)^2}{4\Delta^2} \right]\ket{n},
\label{Gauss}
\ee
where $C$ is for normalization. For such states, $\ex{N}\approx N_0$ and $\va{N}\approx \Delta^2.$
In the inset of Fig.~1, the dashed line corresponds to states of the form Eq.~(\ref{Gauss}) while the solid line
corresponds to points saturating Eq.~(\ref{unc3}). It can be seen that states of the form Eq.~(\ref{Gauss})
are close to saturating Eq.~(\ref{unc3}).
In Fig.~2, we plotted the relative difference between the left hand side and the right hand side
of Eq.~(\ref{unc3}) for particular values of $\Delta$ and $N_0.$

Finally, Fig.~3 shows the distance in the $\va{a}$--$\va{N}$ plane of the points corresponding
to states of the form Eq.~(\ref{Gauss}) from the curve corresponding to states that saturate
Eq.~(\ref{unc3}) \cite{distance}. The distance does not grow with $N$ and remains smaller than $0.15.$
This means that for large $N,$ for which $\va{a}$ and $\va{N}$ cannot be measured with an accuracy of $0.15,$
the 
states Eq.~(\ref{Gauss}) are practically
the intelligent states of the uncertainty relation Eq.~(\ref{unc3}).

\begin{figure}
\centerline{\epsfxsize=3in\epsffile{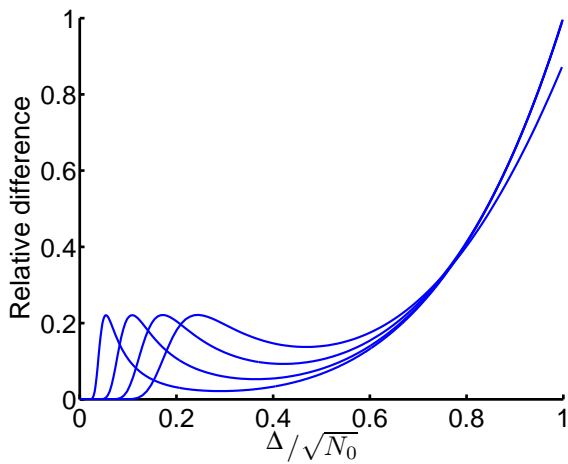}}
\caption{(Color
 online) Numerical test of the inequality Eq.~(\ref{unc3}).
The difference between the left-hand side and the right hand-side of  Eq.~(\ref{unc3}) divided by the right hand side
is shown for states Eq.~(\ref{Gauss}) with a Gaussian wave vector for (from left to right) $N_0=100,25,10$ and $5.$
}
\label{fig_error}
\end{figure}

\begin{figure}
\centerline{\epsfxsize=3.in\epsffile{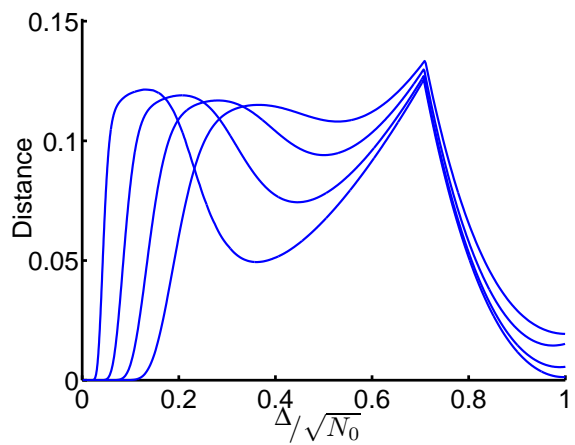}}
\caption{ (Color
 online) Distance of the points corresponding
to states with a Gaussian wave vector Eq.~(\ref{Gauss}) from the curve corresponding to states that saturate
Eq.~(\ref{unc3}) for (from left to right) $N_0=100,25,10$ and $5.$ $\Delta$ and $N_0$ are the parameters
of the state Eq.~(\ref{Gauss}) .
}
\label{fig_error2}
\end{figure}

The states Eq.~(\ref{Gauss}) are the subset of the  number-phase intelligent states
called $\ket{g'}$ presented in Ref.~\cite{VP90}. There, the coefficients of $\ket{n}$
have a Gaussian dependence on $n,$ just as in Eq.~(\ref{Gauss}), however, the phase
of the coefficients is not zero but has a linear dependence on $n.$
As we have already discussed, a state vector with a phase with a linear dependence
on $n$ has the same values for $\va{a}, \va{N}$ and $\ex{N}$ as a state
with zero phase. Thus, all the $\ket{g'}$ states presented in Ref.~\cite{VP90} are
very close to saturating Eq.~(\ref{unc3}).
Hence, our inequality makes it possible to define number-phase intelligent states
and amplitude squeezing \cite{AM87} without a reference to
a phase operator.

There are other states known to be number-phase intelligent states \cite{KY86,D90}.
We now consider the states presented in Ref.~\cite{AJ91}.
They are defined as the superposition of
coherent states on a circle
\be
\ket{\alpha_0,u}\propto \int_{-\infty}^{+\infty} \exp(-\tfrac{1}{2}u^2\phi^2-i\delta\phi)
\ket{\alpha_0 e^{i\phi}}d\phi,\label{intel}
\ee
where $\delta=\alpha_0^2.$ The overlap with Fock states is
\bea
\braket{\alpha_0,u\vert n}&\propto&\frac{\alpha_0^n}{\sqrt{n!}}
\exp\left[-\frac{(n-\delta)^2}{2u^2}\right]\nonumber\\&=&\braket{\alpha_0\vert n}
\times\exp\left[-\frac{(n-\delta)^2}{2u^2}\right].
\eea
The second expression stresses the fact that we have the overlap
of a coherent state $\ket{\alpha_0}$ and a Fock state $\ket{n},$ multiplied
by a Gaussian centered around $\alpha_0^2,$ that is,
the expectation value of the particle number for $\ket{\alpha_0}.$
Thus, $\ket{\alpha_0,u}$ has an almost Gaussian wave vector for large $N$ in the number basis.
Hence, these states give similar results numerically for our uncertainty relation Eq.~(\ref{unc3}) as does Eq.~(\ref{Gauss}).

\subsection{Squeezed coherent states}
It is natural to ask to what extent squeezed coherent states
approach the curve defined by Eq.~(\ref{unc3}).
Squeezed coherent states can be obtained from the vacuum state as \cite{Loudon,HG88}
\be
\ket{\alpha,\zeta}=D(\alpha)S(\zeta)\ket{0},\label{sqcoh}
\ee
where $D$ is the displacement operator and $S$ is the squeezing operator.
Next, we will use the relationships
\bea D^\dagger(\alpha)aD(\alpha)&=&a+\alpha, \nonumber\\
S^\dagger(\zeta)aS(\zeta)&=&a\cosh(s)-a^\dagger e^{i\vartheta}\sinh(s),\label{DS}\eea
where $\zeta=se^{i\vartheta}.$ Hence, with $\alpha=\vert\alpha\vert e^{i\theta},$ we obtain
\bea
\ex{N}_{\ket{\alpha,\zeta}}&=& \sinh^2(s) +\vert\alpha\vert^2,\nonumber\\
\va{a}_{\ket{\alpha,\zeta}}&=&\sinh^2 s,\nonumber\\
\va{N}_{\ket{\alpha,\zeta}}&=&\vert\alpha\vert^2\left[
\cosh(2s)-\sinh(2s)\cos(2\theta-\vartheta)
\right]\nonumber\\&+&2\sinh^2(s) \left[1+\sinh^2(s)\right].\label{dnsc}
\eea

For given $\vert\alpha\vert$ and $s,$ the variance $\va{N}$ in Eq.~(\ref{dnsc}) is minimal if
$\cos(2\theta-\vartheta)=1.$ This is fulfilled, for example, if
$\theta=\vartheta=0,$ that is, both $\zeta$ and $\alpha$ are real and
nonnegative. Hence,
\bea
\va{N}_{\min}(\vert\alpha\vert,s)&=&\vert\alpha\vert^2\left[
\cosh(2s)-\sinh(2s)
\right]\nonumber\\&+&2\sinh^2(s) \left[1+\sinh^2(s)\right].\label{nmin}
\eea
Based on Eq.~(\ref{nmin}), we obtain
 the smallest possible
$\va{N}$
for squeezed coherent states, for given $\va{a}$ and $\ex{N}$ as
\bea
\va{N}_{\min}&=&[\ex{N}-\va{a}]\left[\sqrt{1+\va{a}}
-\sqrt{\va{a}}\right]^2\nonumber\\&+&2\va{a}[1+\va{a}].
\label{scs}
\eea
A dashed curve  in Fig.~1 corresponds to Eq.~(\ref{scs}).

Let us interpret this result.
Since $\va{a}$ and $\va{N}$ are invariant under a rotation around the origin in the $x$--$p$ plane,
we can start from coherent states $\ket{\alpha}$ with a real and positive $\alpha.$
Then, the state we considered for the curve Eq.~(\ref{scs}) corresponds to squeezing
of the $x$ quadrature component, which is called "number-squeezing" in the literature
(e.g., see Ref.~\cite{Loudon})
and it reduces the number variance for a small amount of squeezing.
Thus, for small squeezing Eq.~(\ref{scs}) is not far from the bound given by
Eq.~(\ref{unc3}).
With further squeezing, the number variance starts to grow.
Thus, for large
$\va{a}$ there are no squeezed coherent states giving an almost minimal
particle number variance, and one has to look for non-Gaussian states
for that. As a by-product of our discussion,
note that the non-Gaussianness of quantum states
can be verified by measuring only $\va{a}$ and $\va{N}.$

\subsection{Displaced Fock states}

Displaced Fock states are defined as \cite{W91}
\be
\ket{\alpha,n}=D(\alpha)\ket{n}. \label{df}
\ee
Using Eq.~(\ref{DS}), we obtain
\bea
\ex{N}_{\ket{\alpha,n}}&=&n+\vert\alpha\vert^2,\nonumber\\
\va{N}_{\ket{\alpha,n}}&=&(2n+1)\vert\alpha\vert^2,\nonumber\\
\va{a}_{\ket{\alpha,n}}&=&n.
\eea
Hence, for displaced Fock states  we get the equation
\be
\va{N}=\left[2\va{a}+1\right]\left[\ex{N}-\va{a}\right], \label{dF}
\ee
where $\va{a}$ must be a non-negative integer.
It is fulfilled by both Fock states and coherent states.
Other points in the $\va{a}-\va{N}$ plane satisfying
Eq.~(\ref{dF}) are very far from saturating
Eq.~(\ref{unc3}).

\subsection{Photon-added coherent states}

Photon added coherent states are defined as \cite{AT91}
\be
\ket{\alpha,m}\propto (a^\dagger)^m \ket{\alpha}.\label{pac}
\ee
They are close to saturating Eq.~(\ref{unc3}),
as can be seen in Fig.~1.

\subsection{Eigenstates of $a^\dagger a + {\rm const.}\times a$}

According to Heisenberg's method,
states that minimize the uncertainty product $\va{X}\va{Y}$
for Hermitian $X$ and $Y$ with a constant commutator
are the eigenstates of
$X+icY,$ where $c$ is some constant \cite{J68}.
While in Eq.~(\ref{unc3})
we do not have the product of the uncertainties of two Hermitian observables,
this method can still give us the idea of
considering the states $\ket{d,k}$ defined through the eigenvalue equation
\be
(a^\dagger a + d a)\ket{d,k}=k\ket{d,k},
\ee
where $d$ and $k$ are constants.
Writing $\ket{d,k}$ as $\sum_k c_n' \ket{n},$  we obtain
\be
 c_{n+1}'=\frac{(k-n)}{d \sqrt{n+1}}c_n' \label{d}
\ee
for the coefficients.
Equation~(\ref{d}) leads to a normalizable wave vector only if $k$ is a non-negative integer.
In this case, $c_l=0$ for all $l>k.$
Numerical evidence suggests that states $\ket{d,k}$ are close
to saturating Eq.~(\ref{unc3}), but they are inferior
to the states given by Eq.~(\ref{Gauss}) .

\subsection{States minimizing $\va{N}$ for given $\va{a}$ and  $\ex{N}$}

Let us now look for states that
minimize $\va{N}$ for given $\va{a}$ and  $\ex{N}.$
For that, we will follow an approach similar to the one presented in Ref.~\cite{J68}.
Since $\va{a}=\ex{N}-\tfrac{1}{2}(\ex{x}^2+\ex{p}^2),$ this task can be reformulated
as a search for  the states that
minimize $\va{N}$ for given $\ex{x},$ $\ex{p},$ and  $\ex{N}.$
Let us write the state as $\ket{\Phi}=\sum_k c_n'' \ket{n}.$ Hence, we have to look for
the minimum of the function
\bea
&&f(\bra{\Phi},\ket{\Phi},\lambda_N,\lambda_p,\lambda_x)=
\ex{N^2}-N_0^2\nonumber\\&&\;\;\;\;\;\;+\lambda_x(\ex{x}-x_0)
+\lambda_p(\ex{p}-p_0)+\lambda_N(\ex{N}-N_0),\nonumber\\\label{eigen}
\eea
where $\lambda_k$ are Lagrange multipliers. Note that
we do not include
explicitly the $\langle \Phi \vert \Phi \rangle=1$ condition in the function $f.$
The minimum is given
by one of the critical points for which all derivatives are zero.
Equation~(\ref{eigen}) can be rewritten as
\bea
&&f(\bra{\Phi},\ket{\Phi},\lambda_N,\lambda_p,\lambda_x)=
\ex{O(\lambda_N,\lambda_p,\lambda_x)}_{\ket{\Phi}},\nonumber\\\label{fo}
\eea
where $O$ is defined as
\begin{eqnarray}
&&O(\lambda_N,\lambda_p,\lambda_x)\nonumber\\
&&\;\;\;=\lambda_N a^\dagger a + (a^\dagger a)^2+ \left(\frac{\lambda_x+i\lambda_p}{\sqrt{2}}\right) a +  \left(\frac{\lambda_x-i\lambda_p}{\sqrt{2}}\right) a^\dagger.\nonumber\\\label{opNa}
\end{eqnarray}
We have to look for
$\{\ket{\Phi^{(k)}},\lambda_N^{(k)},\lambda_x^{(k)},\lambda_p^{(k)}\}$ that
minimize Eq.~(\ref{fo}).  It is easy to see that $\ket{\Phi^{(k)}}$
must minimize $\ex{O(\lambda_N^{(k)},\lambda_p^{(k)},\lambda_x^{(k)})}.$
Hence, states $\ket{\Phi^{(k)}}$ must be the eigenstates of the operator
$O(\lambda_N^{(k)},\lambda_p^{(k)},\lambda_x^{(k)})$
with the smallest eigenvalue (i.e., they have to be "ground states").
Note that the operator given in Eq.~(\ref{opNa}), appears as a system Hamiltonian in self-consistent calculations
for the Bose-Hubbard model based on the Gutzwiller ansatz \cite{Gutzwiller,priv}.

\section {Discussion}

Let us discuss the use of quantum states that minimize $\va{N}$
for given $\ex{N}$ and $\va{a}.$ They present a trade-off between two
requirements: the smallest possible variance of a randomly
chosen quadrature component and the smallest possible
particle number variance. In a sense, they are similar to
states minimizing $\va{N}$
for given $\va{\phi}.$ The latter
present a trade-off between the smallest possible variances for phase measurements
and for particle number measurements.

Clearly, the right hand side of Eq.~(\ref{unc3}) is not a constant,
but it is a quantity that can be controlled easily in many systems.
Moreover, note that the measurement of $\va{a}$ does not require
the measurement of variances of $x$ and $p$ if we use
$\va{a}=\ex{N}-|\ex{a}|^2=
\ex{N}-\frac{1}{2}(\ex{x}^2+\ex{p}^2).$

A single trapped ion seems to be a good candidate
for testing our inequalities
and realizing quantum states that saturate them \cite{LB03,GZ05,GK09}.
For a trapped ion, $x$ and $p$ are the physical position and momentum coordinates,
and  $N$ determines the energy of the ion.

The uncertainty relation Eq.~(\ref{unc3})
can also be verified experimentally
in a single mode electromagnetic field.
The two orthogonal quadrature components can be measured
for example with homodyne detection \cite{qoptbook}.
The result is not influenced by which two orthogonal components
we choose to measure.

Bose-Einstein condensates of alkali atoms seem to be also
a possible candidate for experiments \cite{DG99,GZ05}.
It is usual to talk about number squeezing in multi-well Bose-Einstein condensates
in the sense that an increase in the the barrier hight between the wells decreases the
number fluctuation within the wells \cite{GF06}.
Here, one has to note that
for cold atoms the particle number is conserved.
Because of that, for a single bosonic mode of cold atoms,
superpositions of
states with different particle numbers are not allowed.
For this reason, for pure states $\va{N}=0.$
It is possible to mix states with different particle numbers
making $\va{N}>0.$ However,  even for such states $\ex{a}=0$ and $\va{a}=\ex{N},$
which makes  our inequalities trivial in such systems.

While we cannot create particles in a single mode, we
can move particles from one mode to another one.
Thus, it is instructive to consider two-mode systems of cold atoms.
The two modes can be realized
with atoms in a double well or
with a single Bose-Einstein condensate of two
state atoms. Let us denote the annihilation operators of the two modes by $a_1$ and $a_2,$ respectively.
The corresponding particle numbers are $N_1$ and $N_2.$
 If $\ex{N_1} \ll \ex{N_2}$ and $\va{N_2}\ll\ex{N_2}^2$ then with the substitution
\bea a &\rightarrow& \frac{a_1a_2^\dagger}{\sqrt{\ex{N_2}}},\nonumber\\ N &\rightarrow& N_1,\eea
the uncertainty relations Eqs.~(\ref{unc}) and (\ref{unc3}) can be tested.
In the Appendix we present relations that do not
require such approximations.

Finally, in the statistical physics of bosonic systems,
$\ex{\Psi(x)}$, i.e., the expectation value of the field operator
plays the role of the order parameter. In this context,
our findings present a quantitative relationship between the
variance of the field operator and the variance of the particle density
$\Psi(x)^\dagger\Psi(x)$.

\section {Summary}

We constructed uncertainty relations with the particle number and the
annihilation operator. The variance of the latter describes
the uncertainty in the phase space,
and is independent of the absolute phase of
the quadrature components. We proposed quantum optical systems in
which our inequality could be tested.

\section*{Acknowledgments}

We thank I.L.~Egusquiza, J.J.~Garc\'{\i}a-Ripoll and M.W.~Mitchell for fruitful discussions.
We thank P.~Adam for interesting discussions on amplitude squeezing.
We thank the support of
the National
Research Fund of Hungary OTKA (Contract No. T049234),
and the Spanish MEC (Ramon y Cajal Programme,
Consolider-Ingenio 2010 project ''QOIT'', project No. FIS2009-12773-C02-02).
I.U.L. acknowledges the support of
a Ph.D. grant of the Basque Government
and thanks for the hospitality of the Department of Theoretical Physics,
where most of the research for this paper
has been completed in a summer project.

\appendix*

\section{Uncertainty relations for two-mode systems}

For the two-mode system, inequalities similar to Eqs.~(\ref{unc}) and (\ref{unc3}) can be found using
the Schwinger representations of the angular momentum operators
\bea
J_l&=&\frac{1}{2}
\left(
\begin{array}{c}
  a_1^\dagger    \\
  a_2 ^\dagger    \\
\end{array}
\right)^T \sigma_l \left(
\begin{array}{c}
  a_1    \\
  a_2     \\
\end{array}
\right),
\eea
for $l=x,y,z$ where $\sigma_l$ are the Pauli spin matrices.
Let us define an operator that is an analog of $a$ in the two mode-system as
\be
\widetilde{a}=J_x-iJ_y\equiv a_1a_2^\dagger.
\ee
With this definition, we have
\bea
(\vert \Delta \widetilde{a} \vert^2)&=&\tfrac{1}{2}[\va{\widetilde{a}}+\va{\widetilde{a}^\dagger}]=\va{J_x}+\va{J_y}, \nonumber\\
\vert \ex{\widetilde{a}} \vert^2 &=&\ex{J_x}^2+\ex{J_y}^2. \label{AAA}
\eea
Using Eq.~(\ref{AAA}) and the Heisenberg uncertainty relation $\va{J_k}\va{J_l}\ge\tfrac{1}{4}\vert\ex{J_m}\vert^2,$
we obtain the analog of Eq.~(\ref{unc})
\be
\va{N_1}(\vert \Delta \widetilde{a} \vert^2)\ge\frac{1}{4}\vert \ex{\widetilde{a}} \vert^2.\label{Jxyz}
\ee
Adding $\tfrac{1}{4}(\vert \Delta \widetilde{a} \vert^2)$ to both sides of Eq.~(\ref{Jxyz}) and using
\be
J_x^2+J_y^2=\frac{1}{2}(N_1+1)(N_2+1)-\frac{1}{2},
\ee
we obtain an analogue of Eq.~(\ref{unc3})
\bea
\left[\va{N_1}+\frac{1}{4}\right](\vert \Delta \widetilde{a} \vert^2) \ge
\frac{1}{8} \left\langle\left(N_1+1\right)\left(N_2+1\right)\right\rangle-\frac{1}{8}.\nonumber\\
\eea
As we mentioned previously, number squeezing with Bose-Einstein condensates in a double-well
can occur if the barrier between the wells increases \cite{GF06}.
Equation.~(\ref{unc3}) bounds the number variance of a well in such systems \cite{DC00}.



\begin{thebibliography}{99}

\bibitem{C68} P. Carruthers and M.M. Nieto, Rev. Mod. Phys. {\bf 40},
411 (1968).
\bibitem{L95} R. Lynch, Phys. Rep. {\bf 256}, 367 (1995).
\bibitem{BR97} S.M. Barnett and P.M. Radmore, {\it Methods in
Theoretical Quantum Optics} (Oxford University Press, Oxford, 1997).
\bibitem{D27} P.A.M. Dirac, Proc. R. Soc. A {\bf 114}, 243 (1927).
\bibitem{SG64} L. Susskind and J. Glogower, Physics {\bf 1}, 49 (1964).
\bibitem{L76} J.-M. L\'evy-Leblond, Ann. Phys. {\bf 101}, 319 (1976).
\bibitem{PB88} D.T. Pegg and S.M. Barnett, Europhys. Lett. {\bf 6}, 483 (1988);
S.M. Barnett and D.T. Pegg, J. Mod. Opt. {\bf 36}, 7 (1989);
D.T. Pegg and S.M. Barnett, \pra {\bf 39}, 1665 (1989).

\bibitem{VP90} J.A. Vaccaro and D.T. Pegg, J. Mod. Opt. {\bf 37} 17 (1990).

\bibitem{AM87} G. D'Ariano,
S. Morosi,
M. Rasetti,
J. Katriel and
A. I. Solomon, Phys. Rev. D {\bf 36}, 2399 (1987).

\bibitem{TC03} G. T\'oth, C. Simon, and J. I. Cirac,
Phys. Rev. A \textbf{68}, 062310 (2003); see also C. Simon, Phys. Rev. A \textbf{66}, 052323 (2002).
\bibitem{entag} For reviews on quantum entanglement and entanglement criteria, see
R. Horodecki,
P. Horodecki, M. Horodecki, and K. Horodecki,
Rev. Mod. Phys. {\bf 81}, 865 (2009);
O. G\"uhne and G. T\'oth, Phys. Rep. {\bf 474}, 1 (2009).
\bibitem{uncert} For entanglement detection based on local
uncertainty relations, see H.F. Hofmann and S. Takeuchi, Phys. Rev. A {\bf 68}, 032103
(2003); O. G\"uhne, Phys. Rev. Lett. {\bf 92}, 117903 (2004).
\bibitem{entdet} For entanglement detection with non-Hermitian operators, see
also M. Hillery, H.T. Dung, and J. Niset, Phys. Rev. A {\bf 80}, 052335 (2009);
M. Hillery and M.S. Zubairy, Phys. Rev. Lett. \textbf{96}, 050503 (2006);
E. Shchukin and W. Vogel, Phys. Rev. Lett. \textbf{95}, 230502 (2005).
\bibitem{corrmat} For entanglement conditions with such moments see L.-M. Duan, G. Giedke, J.I. Cirac, and P. Zoller, Phys. Rev.
Lett. {\bf 84}, 2722 (2000); R. Simon, Phys. Rev. Lett. {\bf 84}, 2726 (2000).
\bibitem{H87} There have been other uncertainty relations in the literature with a bound linear in $\ex{N}.$
See M. Hillery, Phys. Rev. A {\bf 36}, 3796 (1987).

\bibitem{remark} One could also use an alternative definition of the
variance of a non-Hermitian operator 
$
(\vert\Delta a\vert^2)=\tfrac{1}{2}(\ex{a^\dagger a}+\ex{a a^\dagger})-\ex{a}\ex{a^\dagger}.
$
See, for example,
C. M. Caves and B.L. Schumaker, Phys. Rev. A {\bf 31}, 3068 (1985);
B.L. Hu, G. Kang, and A. Matacz, Int. J. Mod. Phys. A {\bf 9}, 991 (1994).

\bibitem{KO99} T.B.L. Kist, M. Orszag, T.A. Brun, and L. Davidovich,
J. Opt. B: Quantum Semiclass. Opt. {\bf 1}, 251 (1999).
\bibitem{SM06} L. Sanza, M.H.Y. Moussa, and K. Furuya, Ann. Phys.
{\bf 321},1206 (2006).
\bibitem{SG93} Y. Salama and N. Gisin, Phys. Lett. A {\bf 181}, 269 (1993).

\bibitem{HK94} B.-L. Hu, G. Kang, and A. Matacz, Int. J. Mod. Phys. A {\bf 9}, 991 (1994).

\bibitem{S86} B. Schumaker, Phys. Rep. {\bf 135}, 317 (1986).

\bibitem{GZ04} C.W. Gardiner and P.  Zoller,  {\it Quantum Noise}, Springer, Berlin, 2004.

\bibitem{distance} If the closest point of the curve had $\va{a}<0,$ we considered the distance from the vertical axis instead.
The distance between points $A$ and $B$ was computed as $\sqrt{[\va{a}_A-\va{a}_B]^2+[\va{N}_A-\va{N}_B]^2}.$

\bibitem{KY86} M. Kitagawa and Y. Yamamoto, \pra {\bf 34} , 3974 (1986).
\bibitem{D90} G.M. D'Ariano, \pra {\bf 41}, 2636 (1990).
\bibitem{AJ91} J. Janszky and An.V. Vinogradov, Phys. Rev. Lett.  {\bf 64}, 2771 (1990);
P. Adam and J. Janszky, Phys. Lett. A {\bf 160}, 506 (1991);
 M. Orszag, R. Ramirez, J.C. Retamal and C. Saavedra, Phys. Rev. Lett. {\bf 68}, 3815 (1992);
 J. Janszky, P. Adam and An.V. Vinogradov, Phys. Rev. Lett. {\bf 68}, 3816 (1992);
 J. Janszky, P. Adam, M. Bertolotti and C. Sibilia, Quantum Opt. {\bf 4}, 163 (1992);
J. Janszky, P. Domokos, and P. Adam, Phys. Rev. A {\bf 48}, 2213 (1993).

\bibitem{HG88} R.W. Henry and S.C. Glozer, Am. J. Phys. {\bf 56}, 318 (1988).
\bibitem{Loudon} R. Loudon,   {\it The Quantum Theory of Light} (Oxford University Press, Oxford, 1973).

\bibitem{W91} 
P. Kr\'al, J. Mod. Opt. {\bf 37}, 889 (1990);
A. W\"unsche, Quantum Opt. {\bf 3}, 359 (1991).

\bibitem{AT91} G.S. Agarwal and K. Tara, Phys. Rev. A {\bf 43}, 492 (1991);
S. Sivakumar,  J. Phys. A {\bf 32} 3441(1999);
A. Zavatta, S. Viciani and M. Bellini, Science
{\bf 306},  660 (2004).

\bibitem{J68} R. Jackiw, J. Math. Phys. {\bf 9}, 339 (1968).

\bibitem{Gutzwiller} M.P.A. Fisher, P.B. Weichman,
G. Grinstein, and D.S. Fisher,
Phys. Rev. B {\bf 40}, 546 (1989); K.
Sheshadri et al., Europhys. Lett. {\bf 22}, 257 (1993); J.K.
Freericks and H. Monien, Europhys. Lett. {\bf 26}, 545 (1994);
L. Amico and V. Penna, Phys. Rev. Lett. {\bf 80}, 2189 (1998);
D. Jaksch, C. Bruder, J.I. Cirac, C.W. Gardiner, and P. Zoller, \prl {\bf 81}, 3108 (1998).

\bibitem{priv} J.J. Garc\'{\i}a-Ripoll, private communication (2009).

\bibitem{LB03} D. Leibfried, R. Blatt, C. Monroe, D. Wineland,  Rev. Mod. Phys. {\bf 75}, 281 (2003).

\bibitem{GZ05} J.J. Garc\'{\i}a-Ripoll, P. Zoller, and J.I. Cirac,
J. Phys. B: At. Mol. Opt. Phys. {\bf 38}, S567 (2005).

\bibitem{GK09} For recent experiments with a single ion, showing
the measurement of $x$ and $p,$ see R. Gerritsma, G. Kirchmair, F. Z\"ahringer, E. Solano, R. Blatt,
and C. F. Roos, Nature {\bf 463}, 68 (2010);
F. Z\"ahringer, G. Kirchmair, R. Gerritsma, E. Solano, R. Blatt, and C. F. Roos,
Phys. Rev. Lett. {\bf 104}, 100503 (2010).

\bibitem{qoptbook} M.O. Scully and M.S. Zubairy, {\it Quantum Optics} 
(Cambridge University Press, Cambridge, 1997).

\bibitem{DG99}  F. Dalfovo, S. Giorgini, L.P. Pitaevskii, and S. Stringari,
Rev. Mod. Phys. {\bf 71}, 463 (1999).

\bibitem{GF06} F. Gerbier, S. F\"olling, A. Widera, O. Mandel, and I. Bloch,
\prl {\bf 96}, 090401 (2006).

\bibitem{DC00} If  $\ex{N_1} \ll \ex{N_2}$ then $\vec{J}=(\ex{J_x},\ex{J_y},\ex{J_z})$ points almost in the $-z$ direction.
It is well known that such systems can be mapped to a harmonic oscillator.
In this case, $J_x$ and $J_y$ play the roles of $x$ and $p,$ respectively.
See L.-M.~Duan, J.I.~Cirac, P.~Zoller, and E.S.~Polzik,
Phys. Rev. Lett. {\bf 85}, 5643 (2000).

\end{thebibliography}
\end{document}